\documentclass[aps,prx,twocolumn,superscriptaddress,longbibliography]{revtex4-1}

\usepackage{amsmath}
\usepackage{amssymb}
\usepackage{graphicx}
\usepackage{afterpage}
\usepackage{amsfonts}
\usepackage{amssymb}
\usepackage{graphicx}

\usepackage{textcomp}

\begin{document}
\title{Systematic study of the influence of coherent phonon wave packets on the lasing properties of a quantum dot ensemble}

\author{D.~Wigger}
\email{d.wigger@wwu.de}
\affiliation{Institut f\"{u}r Festk\"{o}rpertheorie, Universit\"{a}t M\"{u}nster, 48149 M\"{u}nster, Germany}
\author{T. Czerniuk}
\affiliation{Experimentelle Physik 2, Technische Universit\"at Dortmund, 44221 Dortmund, Germany}
\author{D.~E.~Reiter}
\affiliation{Institut f\"{u}r Festk\"{o}rpertheorie, Universit\"{a}t M\"{u}nster, 48149 M\"{u}nster, Germany}
\author{M.~Bayer}
\affiliation{Experimentelle Physik 2, Technische Universit\"at Dortmund, 44221 Dortmund, Germany}
\author{T.~Kuhn}
\affiliation{Institut f\"{u}r Festk\"{o}rpertheorie, Universit\"{a}t M\"{u}nster, 48149 M\"{u}nster, Germany}

\date{\today}


%
\begin{abstract}
Coherent phonons can greatly vary light-matter interaction in semiconductor
nanostructures placed inside an optical resonator on a picosecond time scale.
For an ensemble of quantum dots (QDs) as active laser medium phonons are able to
induce a large enhancement or attenuation of the emission intensity, as has
been recently demonstrated. The physics of this coupled phonon-exciton-light system consists of various effects, which in the experiment typically cannot
be clearly separated, in particular because due to the complicated sample structure a rather complex strain pulse
impinges on the QD ensemble. Here we present a comprehensive
theoretical study how the laser emission is affected by phonon pulses of
various shapes as well as by ensembles with different spectral distributions
of the QDs. This gives insight into the fundamental interaction
dynamics of the coupled phonon-exciton-light system, while it allows us to
clearly discriminate between two prominent effects: the adiabatic shifting of
the ensemble and the shaking effect. This paves the way to a tailored laser
emission controlled by phonons.
\end{abstract}
%


\maketitle

\section{Introduction}
\begin{figure}
\centering
\includegraphics[width=0.8\columnwidth]{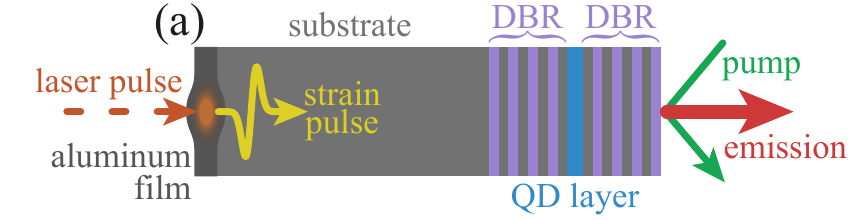}
\includegraphics[width=0.8\columnwidth]{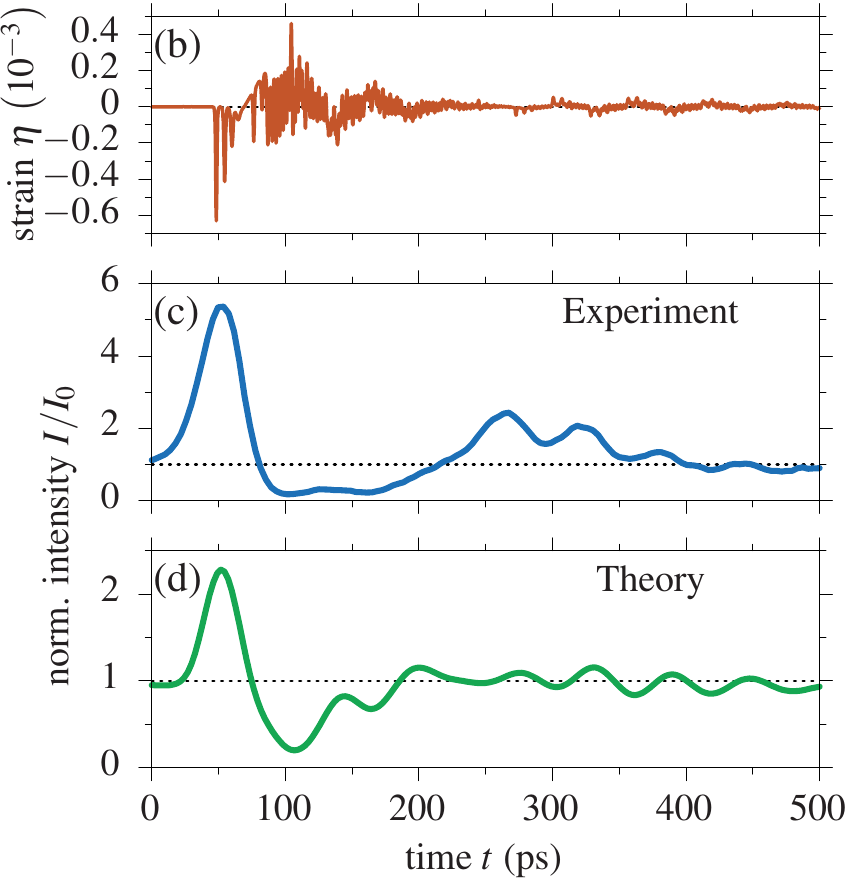}
\caption{(a) Schematic picture of the experiment. (b) Simulated strain field in the experiment. (c) Measured laser intensity normalized to the stationary state intensity. (d) Simulated laser intensity corresponding to (c).
\label{fig:motiv}
}
\end{figure}

At the heart of the growing field of phononics lies the active use of lattice
vibrations in micro- or nanoscale solid state materials. Many applications,
ranging from the excitation of confined phonon modes in optomechanical
systems~\cite{cleland2009opt,aspelmeyer2014cav,volz2016nan} to the control of confined few-level systems by
traveling surface acoustic waves~\cite{couto2009pho,metcalfe2010res,gustafsson2014pro,schulein2015fou,golter2016cou}, prove that phonons
have a great potential to drive and control various solid state systems. A
key optical application of solid state structures is the semiconductor laser.
Here semiconductor quantum dots (QDs) play an increasing role as active laser
medium and, in the extreme limit, even a single QD has been shown to give
rise to lasing \cite{michler2003sin,strauf2011sin,chow2014emi}. While
carrier-phonon interaction is an important ingredient for energy relaxation,
e.g., when bringing carriers from the wetting layer into the laser-active
QDs, for the laser process itself the coupling to phonons often does not play
a crucial role. Only recently, the two fields -- phononics and semiconductor
laser physics -- have been combined by using traveling coherent phonon wave
packets to control the laser
output~\cite{bruggemann2012las,czerniuk2014las,czerniuk2015imp}. We want to briefly recap these findings: In such an experiment, which is schematically shown in
Fig.~\ref{fig:motiv}(a), a layer of QDs, surrounded by two distributed Bragg
reflectors (DBRs), is used as the active laser medium, which can be pumped
optically~\cite{bruggemann2012las,czerniuk2014las} or
electrically~\cite{czerniuk2015imp}. By rapidly heating an aluminum film a
coherent phonon wave packet is created, which travels through the structure.
Due to nonlinear propagation in the material and the transit through the Bragg
mirror the strain profile arriving at the QD layer, shown in
Fig.~\ref{fig:motiv}(b), is rather complex. When the phonons impinge on the
QDs the strain shifts the frequency of the optical transition in the QDs. As a result, the laser output can be greatly enhanced, but also quenching can be
induced. This is demonstrated in Fig.~\ref{fig:motiv}(c), where an example of
the measured laser intensity $I$ normalized to the stationary state intensity
$I_0$ is shown. A detailed interpretation of the coupled phonon-exciton-light system is rather sophisticated because depending on the spectral properties of the QD ensemble and the time scale associated with the phonon dynamics different phenomena contribute to the observed modulation of the laser output.

We here develop and investigate a semiclassical laser model, as depicted in
Fig.~\ref{fig:scheme}, to describe the dynamics of the coupled system
consisting of coherent phonon wave packets, QD excitons and laser light. Our
model has already been shown to well reproduce the experimentally observed
laser outputs ~\cite{czerniuk2017pic}. In this paper we now use the model to
gain a deeper understanding of the coupled phonon-exciton-light system by
systematically studying different ensemble and strain scenarios. Due to the
complex strain pulses present in the experiment, a clear interpretation and
in particular a separation between different phenomena is still challenging.
Two effects turned out to be important, when controlling the laser output: On
the one hand the \emph{adiabatic shift} of the ensemble, which dynamically
changes the number of QDs in resonance with the laser cavity. On the other
hand the \emph{shaking effect} of the ensemble, which rapidly brings
initially detuned and therefore highly occupied QD excitons into resonance
with the cavity. In this paper, we are going to use this model to
systematically study the influence of the phonons on the lasing activity and
to discriminate between the phenomena of shifting and shaking. For this
purpose, we will use idealized ensemble shapes (e.g., a flat distribution of
QD transition energies) and idealized strain pulses (e.g., bipolar or
monopolar pulses or monochromatic phonons) which will allow us to separate
the shaking effect from the adiabatic shift in an intuitive way.

By providing a thorough understanding of the coupled phonon-exciton-light
system, we are not only able to understand the complex laser emission seen in
the experiment, but we also lay the foundation to optimize the system, which
will then stimulate further studies on phonon control of solid state lasers
and pave the way for designing applications.

The paper is structured as follows: In Sec.~\ref{sec:theory} the basics of
the semiclassical laser model are summarized. The results are given in
Sec.~\ref{sec:results}, first for a rectangular QD distribution in part
~\ref{sec:rect} and second for a Gaussian ensemble in part~\ref{sec:gauss}.
In Sec.~\ref{sec:compare} we come back to the experimental results in
Fig.~\ref{fig:motiv} and provide a detailed interpretation on the basis of
the results for the idealized cases. The paper closes with some concluding
remarks in Sec~\ref{sec:concl}.

\section{Theoretical model}
\label{sec:theory}
To model the QD laser system we employ a semiclassical laser model in the
usual dipole, rotating wave and slowly varying amplitude
approximation~\cite{haken}. The QDs are taken to be in the strong confinement
limit, such that for the coupling to the laser mode and the coherent phonons
they can be reduced to two-level systems consisting of  ground state
$\left|g\right>_{\rm i}$ and exciton state $\left|x\right>_{\rm i}$ with
transition frequencies $\omega_{\rm i}$, ${\rm i}\in \{1, 2, \dots ,N_{\rm
QD}\}$ labelling the individual QD and $N_{\rm QD}$ being the number of QDs
in the ensemble. The QD is coupled via the dipole matrix element ${\bf
M}_{\rm i}$ to the light field, treated as a classical field with positive
(negative) frequency component ${\bf E}$ (${\bf E}^\ast$) as well as to the
acoustic phonons, described by the phonon creation (annihilation) operators
$b^\dagger_{\bf q}$ ($b^{}_{\bf q}$), ${\bf q}$ denoting the phonon wave
vector. The Hamiltonian of the QD ensemble then reads
\begin{subequations}
\begin{equation}
H = \sum_{\rm i=1}^{N_{\rm QD}} h_{\rm i} \,,\\
\end{equation}
with
\begin{eqnarray}
h_i &=&
\hbar \omega_{\rm i}\left|x\right>_{\rm i} \left<x\right|_{\rm i}
- \left( {\bf M}_{\rm i}\cdot {\bf E} \left|x\right>_{\rm i}\left<g\right|_{\rm i}
+ {\bf M}_{\rm i}^\ast\cdot {\bf E}^\ast \left|g\right>_{\rm i}\left<x\right|_{\rm i}\right) \nonumber \\
& & + \sum_{\bf q} \hbar g^{}_{\bf q}\left( b^{ }_{\bf q} + b^\dagger_{\bf q} \right) \left| x\right>_{\rm i} \left< x\right|_{\rm i} \,.
\end{eqnarray}\end{subequations}
and the exciton-phonon coupling matrix element $g_{\bf q}$. From the
Hamiltonian and Heisenberg's equation of motion we obtain the equation of
motion for the microscopic polarization $p_{\rm i}=\left<\left|g\right>_{\rm
i}\left<x\right|_{\rm i}\right>$ as well as the occupations $x_{\rm
i}=\left<\left|x\right>_{\rm i}\left<x\right|_{\rm i}\right>$ and $g_{\rm
i}=\left<\left|g\right>_{\rm i}\left<g\right|_{\rm i}\right>$. After
truncating the electron-phonon correlation expansion on the mean field level,
which is appropriate for the description of the coupling to an external
coherent phonon wave packet, we get
\begin{subequations}
\begin{eqnarray}
\frac{\rm d}{{\rm d}t} p_{\rm i} &=& -i\left[ \omega_{\rm i} + 2\sum_{\bf q}g^{}_{\bf q}{\rm Re}\left(\left<b^{ }_{\bf q}\right>\right)\right]p_{\rm i} \,,\label{eq:pi}\\
&&\hspace{3.05cm} +\frac{i}{\hbar}{\bf M}_{\rm i}\cdot {\bf E}\left( g_{\rm i}-x_{\rm i} \right) \nonumber\\
\frac{\rm d}{{\rm d}t} x_{\rm i} &=& -\frac{\rm d}{{\rm d}t} g_{\rm i} = -\frac{i}{\hbar} \left( {\bf M}^\ast_{\rm i}\cdot {\bf E}^\ast p_{\rm i} - {\bf M}\cdot {\bf E} \,p_{\rm i}^\ast \right)\,.
\end{eqnarray}\end{subequations}
The dominant exciton-phonon coupling in typical InGaAs-based QDs is provided
by the deformation potential coupling. For this mechanism the mean field
contribution can be written as~\cite{krummheuer2005cou}
\begin{equation}
2\sum_{\bf q} g_{\bf q} {\rm Re}\left(\left<b^{ }_{\bf q}\right>\right) = D\, {\rm div} \left( \left< {\bf u} ( {\bf r}_{\rm i} ,t ) \right> \right) = D \eta(t)
\end{equation}
with $\left<{\bf u}({\bf r}_{\rm i},t)\right>$ being the  expectation value
of the lattice displacement at the position ${\bf r}_{\rm i}$ of the QD and
$\eta(t) = {\rm div} \left( \left< {\bf u} ( {\bf r}_{\rm i} ,t ) \right>
\right)$ is the corresponding strain. $D=D_{\rm e} - D_{\rm h}$ is the deformation
potential of the exciton with $D_{\rm e}$ ($D_{\rm h}$) being the deformation potentials
of electrons and holes. From Eq.~\eqref{eq:pi} we can therefore conclude that
the transition energy of each QD is shifted from the unstrained value
$\hbar\omega_{\rm i}$ to a time-dependent value $\hbar\omega_{\rm x}$, the
shift being proportional to the strain value $\eta(t)$ at time $t$, according
to
\begin{equation}
\hbar\omega_{\rm x}(t) = \hbar\omega_{\rm i} + D\eta(t)\,.
\label{eq:QDL_strainshift}
\end{equation}
By taking the strain field at the position of the QD we have implicitly
assumed that the wavelength of the phonons contributing to the phonon wave
packet is much larger than the size of the QDs, such that the strain
value does not vary over the size of the QD. Typical values for the
deformation potential are in the range of
$D=-10$~eV~\cite{krummheuer2005pur}. Together with the strain amplitudes in
the experiment [see Fig.~\ref{fig:motiv}(a)] energy shifts of a few meV are
reached. In the following we will directly treat the product $D\eta$ as a
strain mediated energy shift. In the current form of the strain control it is
a good assumption to treat the phonons as plane waves hitting the QD layer
perpendicularly, such that at a given time $t$ all QDs in the ensemble
experience the same shift. One could also think of phonon wave packets that
arrive at a certain angle at the QD layer or that travel in-plane through the
QD ensemble, where QDs at different positions would interact with different
parts of the strain wave. For the description of such a scenario a
spatio-temporal extension of the QD laser model would be
needed~\cite{hess1996max1, hess1996max2, gehrig2002mes,gehrig2004dyn}.

Because the exciton transition energies in the QD ensemble are closely
spaced, we introduce the spectral distribution of the ensemble
\begin{equation}
n(\omega) = \sum_{\rm i=1}^{N_{\rm QD}} \delta(\omega-\omega_{\rm i})
\end{equation}
and move to a continuous description labeling the QDs by their transition
frequencies, i.e., $\hbar\omega_{\rm i}\to \hbar\omega$, $(p_{\rm i},g_{\rm
i},x_{\rm i})\to(p_\omega,g_\omega,x_\omega)$, and $\sum_{\rm i}\to\int
n(\omega)\,{\rm d}\omega$. The dynamics of the cavity field ${\bf E}$ is
driven by the macroscopic polarization of the whole ensemble
\begin{equation}
{\bf P} = \sum_{\rm i} {\bf M}_{\rm i}\,p_{\rm i} \approx {\bf M} \int n(\omega)p_\omega\,{\rm d}\omega\,,
\end{equation}
where we have assumed that the variation of the dipole matrix elements over
the QD ensemble is negligible. By separating the electric field ${\bf E}({\bf
r},t)$ into the normalized mode function of the microcavity ${\bf U}({\bf
r})$ and a time-dependent amplitude $E(t)$ according to ${\bf E}({\bf
r},t)=E(t) {\bf U}({\bf r})$ and introducing the dimensionless cavity field
amplitude $\mathcal E(t)$ with $E(t) = i \sqrt{\hbar \omega_{\rm c} / (2 \epsilon_0)}\,
\mathcal E(t)$ we can write
\begin{equation}
\frac{{\bf M} \cdot {\bf E}({\bf r}_{\rm i})}{\hbar} = G \mathcal E
\end{equation}
with the coupling constant
\begin{equation}
G = i\sqrt{\frac{\omega_{\rm c}}{2\hbar \epsilon_0}} {\bf M} \cdot {\bf U}({\bf r}_{\rm i})\,.
\end{equation}
Here, $\omega_c$ denotes the cavity frequency, $\epsilon_0$ is the vacuum
permittivity  and we have assumed that the coupling constant has
approximately the same value at all QDs of the ensemble, which is reasonable
since all QDs are in a thin layer in the center of the cavity.

\begin{figure}
\centering
\includegraphics[width=0.7\columnwidth]{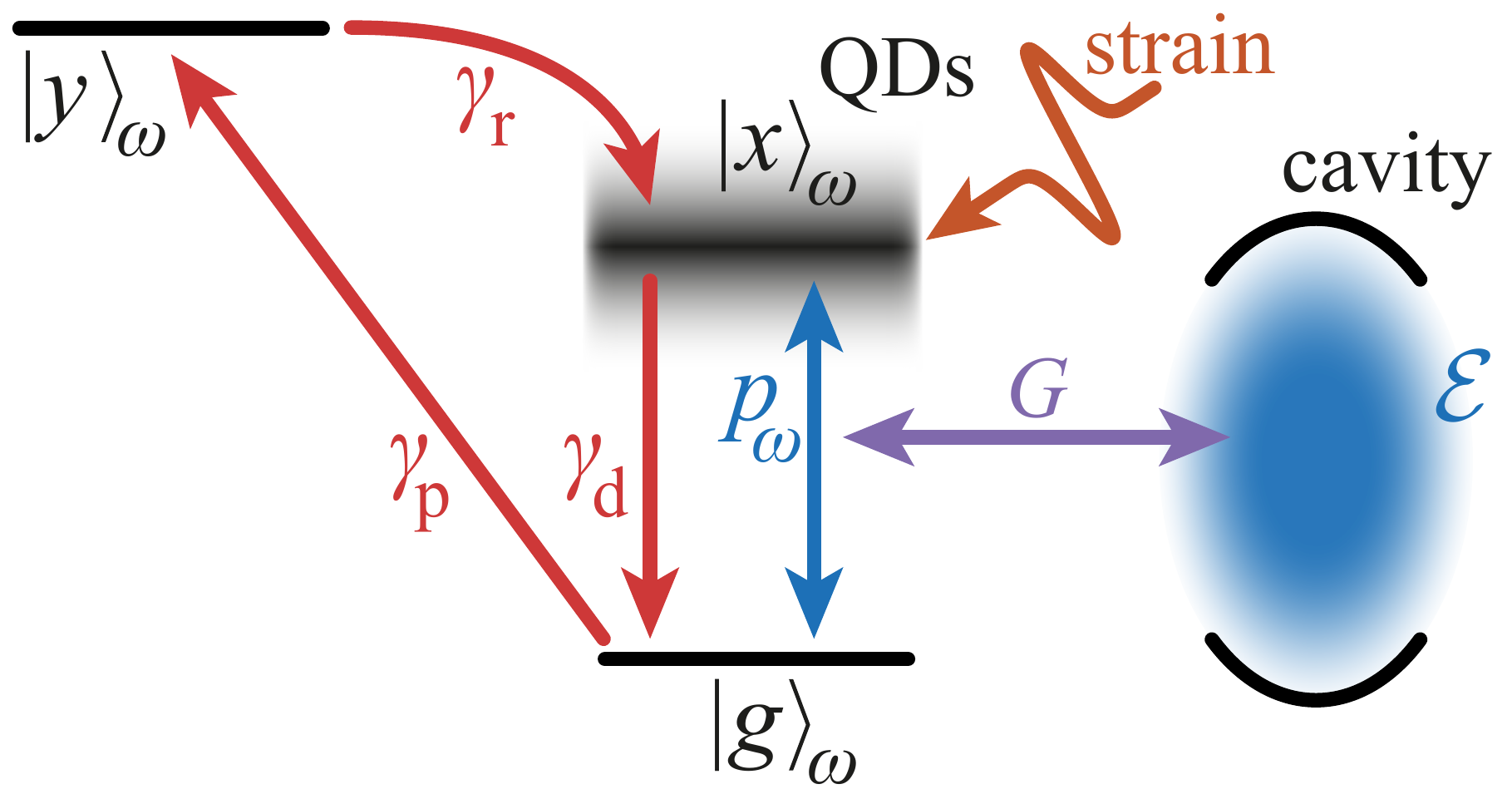}
\caption{Sketch of the theoretical model.
\label{fig:scheme}
}
\end{figure}

To simulate the pumping process we include the occupation of an additional
level $y_\omega = \left<\left|y\right>_\omega\left<y\right|_\omega\right>$,
which can be thought of as wetting layer state and is phenomenologically
coupled to the other levels via constant rates. With this we finally arrive
at the following closed set of equations~\cite{haken}:
\begin{subequations}
\begin{eqnarray}
\frac{\rm d}{{\rm d}t} \mathcal{E} &=& -\gamma_{\rm l} \mathcal{E} - i G^\ast \int n(\omega) p_\omega\,{\rm d}\omega\,,\label{eq:QDL_A}\\
\frac{\rm d}{{\rm d}t} p_\omega &=&  -\frac12 (\gamma_{\rm d}+\gamma_{\rm p}+2\gamma) p_\omega + i \frac{\delta_\omega}{\hbar} p_\omega  \label{eq:QDL_p}\\
&&\hspace{3.05cm} + {\rm i} G \mathcal{E} \left(x_\omega-g_\omega\right) \,, \nonumber \\
\frac{\rm d}{{\rm d}t}  g_\omega &=& -\gamma_{\rm p} g_\omega + \gamma_{\rm d} x_\omega - i \left(G^\ast\mathcal{E}^\ast p_\omega -G\mathcal{E} p_\omega^\ast \right)\,,  \label{eq:QDL_g} \\
\frac{\rm d}{{\rm d}t}  x_\omega &=& -\gamma_{\rm d} x_\omega + \gamma_{\rm r} y_\omega + i \left(G^\ast\mathcal{E}^\ast p_\omega - G\mathcal{E} p_\omega^\ast \right)\,,  \label{eq:QDL_x}\\
\frac{\rm d}{{\rm d}t}  y_\omega &=& -\gamma_{\rm r} y_\omega + \gamma_{\rm p} g_\omega\,,
\end{eqnarray}\label{eq:QDL}\end{subequations}
where $\delta_\omega(t)=\hbar\omega_{\rm c}-\hbar\omega_{\rm x}(t)$ is the
detuning between the cavity mode and the considered exciton transition. Note
that the equations are given in a frame which rotates with the frequency of
the cavity mode. To simulate the vacuum fluctuations and start the lasing
process we add a white noise source to the field amplitude $\mathcal E$ in
every time step. The rates $\gamma_{\rm i}$ in the equations include the pump
rate $\gamma_{\rm p}$ from $\left|g\right>_\omega$ to
$\left|y\right>_\omega$, the relaxation rate $\gamma_{\rm r}$ from
$\left|y\right>_\omega$ to $\left|x\right>_\omega$ and the spontaneous decay
rate $\gamma_{\rm d}$ from $\left|x\right>_\omega$ to $\left|g\right>_\omega$
as shown in Fig.~\ref{fig:scheme}. Additionally $\gamma_{\rm l}$ describes
the cavity loss rate and $\gamma$ a pure dephasing rate, which only acts on
the polarization $p_\omega$. Following the results in
Ref.~\cite{czerniuk2017pic} we choose the same parameters for the laser
system, i.e., $\gamma_{\rm l}=0.4$~ps$^{-1}$, $\gamma_{\rm
d}=0.03$~ps$^{-1}$, $\gamma_{\rm r}=0.5$~ps$^{-1}$, $\gamma=1$~ps$^{-1}$, and
$G=2.8$~ps$^{-1}$. Already in Ref.~\cite{czerniuk2017pic} it was shown that
the strongest laser response on the strain field happens for pump rates that
are slightly above the threshold pump rate, which for our system parameters
is approximately the spontaneous decay rate $\gamma_{\rm d}$. Thus we define
a shifted pump rate as
\begin{equation}
\Gamma_{\rm p} = \gamma_{\rm p}-\gamma_{\rm d}\,.
\end{equation}
%

%
\section{Results}
\label{sec:results}
The main goal for the paper is to separate various effects which contribute
to the modulation of the laser output, in particular the shaking and the
adiabatic shift. To eliminate the latter one, we first consider a QD ensemble
in which the number of dots in resonance with the cavity mode remains
constant when the exciton energy varies. The Gaussian ensemble, will be
discussed in Sec.~\ref{sec:gauss}.

The central quantity to illustrate the dynamics of the QD ensemble is the
inversion density
\begin{equation}
d(\omega)=\left(x_\omega-g_\omega\right)n(\omega)\,,
\end{equation}
which indicates how many QDs with a given transition frequency $\omega$ are
in the excited state. It further reflects the spectral shape of the ensemble.
%
\subsection{Ensemble with rectangular QD distribution}
\label{sec:rect}
To analyze the shaking effect we choose an idealized ensemble shape being
rectangular with the spectral QD density given by
\begin{equation}
n(\omega) =
\left\{
  \begin{array}{c c c}
	\frac{\hbar N_{\rm QD}}{\Delta_{\rm QD}} & {\rm for} & -\frac{\Delta_{\rm QD}}{2}<\hbar\omega-\hbar\omega_0<\frac{\Delta_{\rm QD}}{2} \\
	0 & {\rm otherwise}&
  \end{array}
  \right.\, ,
\end{equation}
where $N_{\rm QD}=5\times10^4$ is the total number of QDs and $\Delta_{\rm
QD}$ is the width around the center $\hbar\omega_0$ of the ensemble. By this
choice the number of QDs in resonance with the laser cavity is always the
same, as long as the cavity resonance stays in the range of the ensemble. A
schematic picture of the ensemble is shown in the inset in
Fig.~\ref{fig:rect_pulse}(a). The cavity resonance is supposed to be
initially in the center of the ensemble. Since the pumping is taken to be the
same for the whole ensemble, an energetic shift of the ensemble will bring
highly occupied QD excitons into resonance with the cavity and therefore
influence the output of the laser. This process is called shaking of the
%
\subsubsection{Single strain pulse}
We start our investigation with a single bipolar strain pulse leading to the
energy shift
\begin{equation}
D\eta(t) = -\eta^{\rm D}_0\, \frac{t}{\tau_{\rm s}} \exp\left\{\frac12 \left[1 - \left(\frac{t}{\tau_{\rm s}}\right)^2\right]\right\}\,,
\label{eq:strain_pulse}
\end{equation}
where the pulse is centred around $t=0$ and has a width of $2\tau_{\rm s}$
between minimum and maximum. In the following, the maximum amplitude of the
energetic shift $\eta_0^{\rm D}$ will for brevity be referred to as strain
amplitude, but note that this always represents the product $D\eta$. The
negative sign in Eq.~\eqref{eq:strain_pulse} reflects the negative sign of the
deformation potential $D$ and has been introduced such that a positive strain
corresponds also to a positive value of $\eta^{\rm D}_0$. The strain dynamics
in Eq.\eqref{eq:strain_pulse} corresponds to a Gaussian phonon wave packet in
the lattice displacement, which is for example generated initially in the
experiment~\cite{scherbakov2007chi} or by an ultrafast optical excitation of
a single QD~\cite{wigger2013flu}.

\begin{figure}[h]
\centering
\includegraphics[width=0.8\columnwidth]{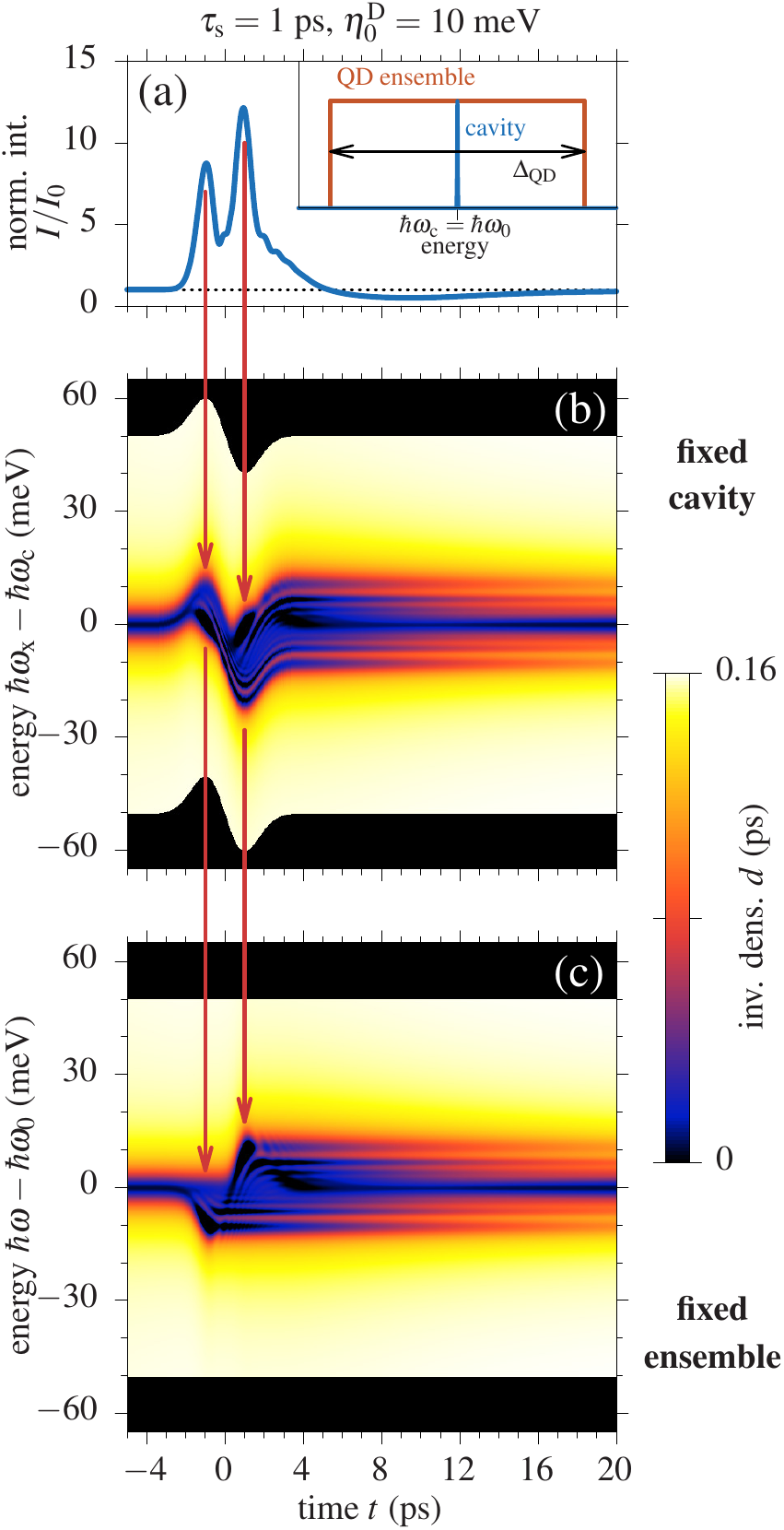}
\caption{Pulsed modulation of a rectangular ensemble. The strain pulse is bipolar with
a duration of $\tau_{\rm s}=1$~ps. (a)~Normalized intensity $I/I_0$ as a function
of time. The inset shows a schematic picture of the ensemble shape. (b), (c)
Inversion density $d$ as a function of time and energy with fixed cavity resonance
in (b) and fixed ensemble position in (c).
\label{fig:rect_pulse}
}
\end{figure}

Figure~\ref{fig:rect_pulse} summarizes the response of the QD laser system on
the interaction with a strain pulse given by Eq.\eqref{eq:strain_pulse} with a
duration of $\tau_{\rm s}=1$~ps and a strain amplitude of $\eta_0^{\rm
D}=10$~meV, the width of the QD ensemble is $\Delta_{\rm QD}=100$ meV. The
strain pulse arrives, when the laser has reached its stationary state. We
choose the pump rate to $\Gamma_{\rm p}=20$~\textmu s$^{-1}$. The laser
intensity $I$ normalized to the intensity without any strain $I_0$ is plotted
in Fig.~\ref{fig:rect_pulse}(a) as a function of time $t$. The laser output
is dominated by two enhancement peaks that reach values of about $I/I_0=10$.
The maxima are at the positions of the extrema of the strain pulse, marked by
red arrows. After the pulse the intensity shows a period of attenuation
around $t=10$~ps and eventually goes back to 1.

To get a better insight into the origin of the variation of the laser output
Fig.~\ref{fig:rect_pulse}(b) and (c) show the inversion density $d$ as a
function of time $t$ and exciton energy $\hbar\omega$. Because in the
equation of motion for the microscopic polarization $p_\omega$
[Eq.~\eqref{eq:QDL_p}] only the detuning between the cavity mode and the
considered QD energy $\delta_\omega$ enters, it is not important for the
simulation if the QD transition energies are shifted by the strain according
to Eq.~\eqref{eq:QDL_strainshift} or if all QD transitions stay at the same
energies and the cavity resonance is shifted by the same amount but in the
opposite energetic direction. Thus it is possible to present the dynamics of
the inversion density of the QD ensemble either with a fixed cavity resonance
energy and a strain affected QD ensemble (i.e., as a function of
$\hbar\omega_{\rm x}-\hbar\omega_{\rm c}$)  or with a fixed ensemble
distribution ($\hbar\omega-\hbar\omega_0$) and a cavity resonance that is
shifted by the strain dynamics. The former is realized in
Fig.~\ref{fig:rect_pulse}(b) and the latter in (c). Both representations are
shown here because in either case different aspects of the inversion dynamics
can be explained more easily. Before the strain pulse, i.e., at
$t\lesssim-4$~ps both inversion density representations look the same. The
region between $-50$ and 50~meV depicts the spectrum of the QD ensemble, for
larger energies there are no QDs which results in a vanishing density
(black). At the edges of the ensemble the QDs are highly occupied, which is
seen by the bright color, but in the center, at the position of the cavity
mode, an energy range with reduced inversion densities appears. This
depression of the exciton occupation represents the QDs that contribute to
the laser process and is called \emph{spectral hole}. During the interaction
with the strain pulse, i.e., for $-4\ {\rm ps}\lesssim t \lesssim 4$~ps, the
actual influence of the strain can be seen. Let us first focus on the
representation in (b). The energetic position of the ensemble is shifted
proportional to the strain amplitude, therefore the complete colored area
shifts to larger energies, reaches a maximum shift of $\eta_0^{\rm D}=10$~meV
at $-1$~ps (left red arrow). It is then shifted toward smaller energies,
reaching again the maximum shift at 1~ps (right red arrow) and goes back to
its initial position. During the whole process the cavity resonance stays at
$\hbar\omega_{\rm x}-\hbar\omega_{\rm c}=0$. In contrast, the spectral hole
does not stay at the energy of the cavity, it rather follows the strain shift
of the ensemble toward positive energies. While this happens the shape of the
spectral hole is strongly modified. After the strain pulse at $t>4$~ps the
final spectral hole consists of multiple minima distributed symmetrically
around the cavity at 0~meV. For increasing times the outer inversion minima
decay and only the original spectral hole remains when the laser comes back
to its steady state.

We now come back to the laser output in (a) and directly compare the
enhancement and attenuation features with the inversion density. The red
arrows mark both the maxima of the laser intensity and the extrema of the
strain shift. At the arrowheads two dark spots appear in (b), which represent
very small or even negative values of $d$. Note that due to clarity the color
scale in Fig.~\ref{fig:rect_pulse}(b) and (c) starts at $d=0$, although
negative values may appear when stimulated emission de-occupies a QD exciton.
The origin of these strong spectral holes can best be explained in the
picture of a fixed ensemble and a strain shifted cavity resonance in (c). The
fact that the edges of the colored area are fixed shows that the QD energies
are unaffected by the strain. Due to the strain shift of the cavity resonance
it first moves to smaller energies and reaches
$\hbar\omega-\hbar\omega_0=-10$~meV at $t=-1$~ps, which is marked by the left
red arrow. We see that at this time the strong deepening of the spectral hole
happens. The reason is, that the cavity moves toward highly occupied QDs
faster than they can fully react on the presence of the cavity mode. When the
cavity shift reaches its maximum at $-1$~ps, the shift is sufficiently slow
for the QDs that are in resonance with the cavity during that time window to
emit stimulatedly into the laser mode. Due to this switch-on process of QDs
that were previously not in resonance with the laser mode, the output
increases drastically resulting in the first peak in (a). The same happens
for the second turning point of the strain shift at $t=1$~ps. After the
strain has shifted the cavity out of resonance again the QDs in these two
strong spectral holes are significantly detuned from the laser mode and can
not contribute to the laser output any more. The re-occupation of the dots
appears as decay of the spectral holes. The last feature of the laser
intensity in (a) is the attenuation between $t=7$ and 16~ps. After the strain
pulse, the cavity resonance is again at the center of the ensemble and the
QDs in the region of the additional outer spectral holes do not contribute to
the laser process any longer. This means that only the QDs within the
spectral hole at $t<-4$~ps, i.e., before the strain pulse, are important for
the intensity. A close look at the shape of the central spectral hole reveals
that it is less deep than before the strain pulse. During the pulse the QD
excitons within the spectral hole were detuned from the cavity resonance and
are therefore higher occupied directly after the pulse. They need a while to
reach the stationary laser state, which has smaller inversions, i.e., a
deeper spectral hole. This also results in the reduced intensity. The
original stationary laser state recovers completely at $t\approx20$~ps.

As a side remark, we note that for strain pulses that are significantly
longer in time ($\tau_{\rm s}\gtrsim 30$~ps) the spectral hole can
adiabatically follow the shift of the ensemble, such that always the same
amount of QDs takes part in the laser process and the intensity stays
constant (not shown).
%
\subsubsection{Harmonic modulation}
\begin{figure*}
\centering
\includegraphics[width=0.8\textwidth]{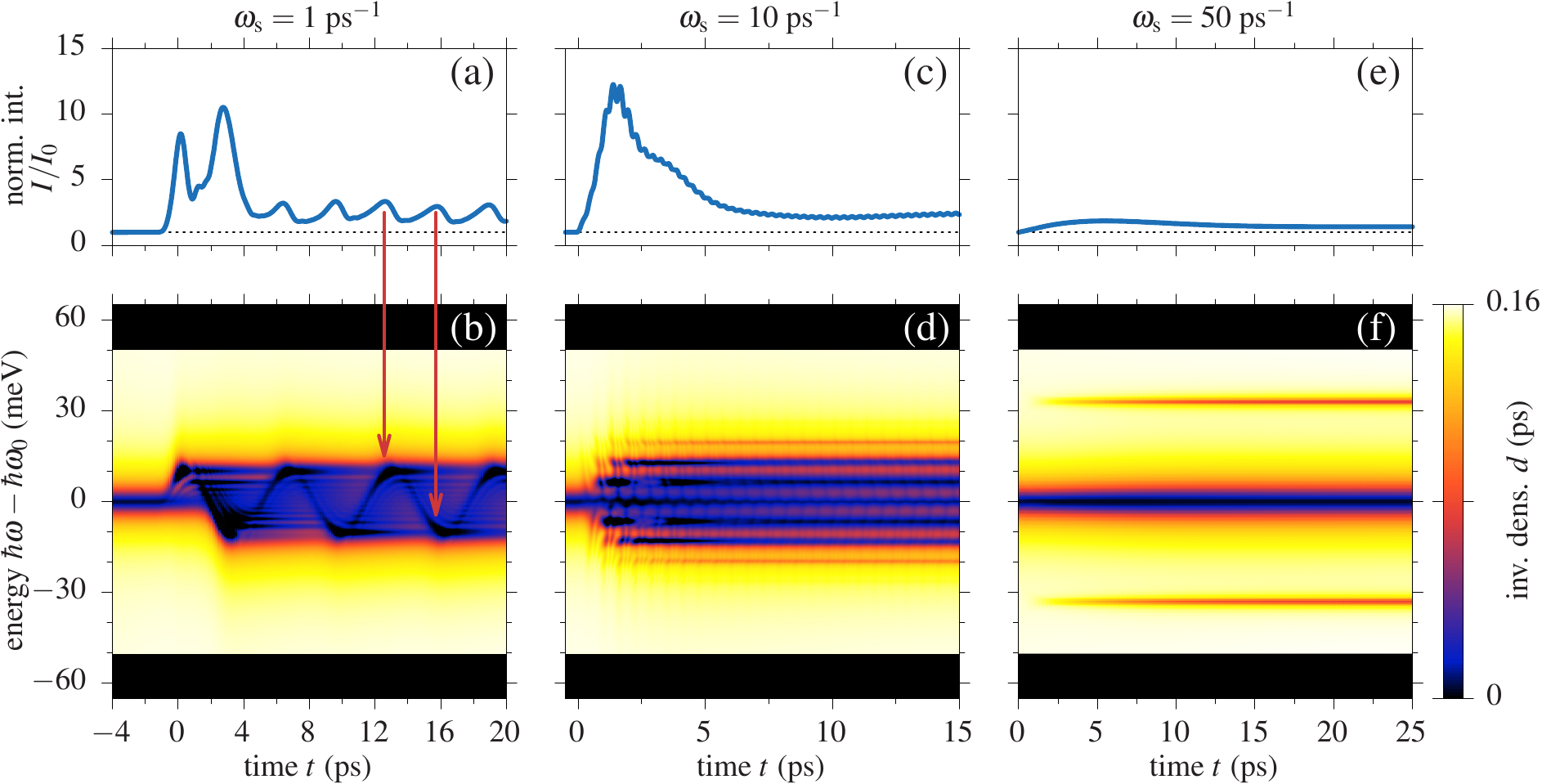}
\caption{Harmonic modulation of a rectangular ensemble. Top row: Normalized
intensity $I/I_0$ as a function of time. Bottom row: Inversion density
$d$ as a function of time and energy with fixed ensemble position. (a), (b)
Strain frequency $\omega_{\rm s}=1$~ps$^{-1}$; (c), (d) $\omega_{\rm s}=10$~ps$^{-1}$; (e),
(f) $\omega_{\rm s}=50$~ps$^{-1}$.
\label{fig:rect_harm}
}
\end{figure*}
We have seen that a shift of the QD ensemble on the picosecond time scale
brings the laser system out of equilibrium and results in variations of the
output intensity on the order of tenfold enhancements. In the second example
we tackle the question, under which conditions a laser system driven by a
harmonic strain perturbation reaches a new stationary state.

Besides the fundamental interest, such a study is well connected to recent
research. Monochromatic coherent phonons can be optically created in
semiconductor superlattices. A recent experiment demonstrates narrow band
phonon generation with frequencies of more than 2~THz~\cite{huynh2008sub,jiang2016aco},
which is comparable to the values studied in the following. To make maximum
use of these vibrations, one can tailor a phononic cavity for the desired
phonon frequency and embed it into the optical resonator -- in this way the
generated phonons become confined in the same place as the active
medium~\cite{trigo2002con}. In passive optomechanical resonators the
localization of phonons and photons results in an enhanced coupling
strength~\cite{fainstein2013str}, while in a laser device the resonant strain
field leads to a harmonic modulation of the lasing
emission~\cite{czerniuk2014las}. We model the harmonic strain field by:

\begin{eqnarray}
D\eta(t) = -\eta^{\rm D}_0 \left\{
  \begin{array}{l c  rcccl}
	0    & ~ &  && t &\!\!<\!\!& -\tau_{\rm s}/4\\
    \cos^2\left(\omega_{\rm s}t\right) & {\rm for} & -\tau_{\rm s}/4 &\!\!<\!\!& t &\!\!<\!\!& 0\\
    \cos\left(\omega_{\rm s}t\right) &  & 0 &\!\!<\!\!& t &&
  \end{array}
  \right. \\
{\rm with}\qquad\tau_{\rm s} = 2\pi/\omega_{\rm s}\,,\nonumber
\label{eq:QDL_strain_harm}
\end{eqnarray}
i.e., the harmonic strain oscillation is smoothly switched on between
$t=-\tau_{\rm s}/4$ and $t=0$, where $\tau_{\rm s}$ is the period of the
strain frequency $\omega_{\rm s}$. The laser has reached its stationary
output long before the strain oscillation is switched on.

In Fig.~\ref{fig:rect_harm} three different strain frequencies are shown for
a pump rate of $\Gamma_{\rm p}=20$~\textmu s$^{-1}$. In the left column [(a),
(b)] we have $\omega_{\rm s}=1$ ps$^{-1}$, in the middle [(c), (d)]
$\omega_{\rm s}=10$ ps$^{-1}$ and on the right [(e), (f)] $\omega_{\rm s}=50$
ps$^{-1}$. The upper row presents the normalized laser intensity $I/I_0$ as a
function of time $t$ and the bottom row the inversion density for a fixed
ensemble as a function of time and exciton energy
$\hbar\omega-\hbar\omega_0$. Note that in this representation the energy of
the cavity resonance is shifted by the strain.

We start the discussion of the harmonically driven laser system with
$\omega_{\rm s}=1$~ps $^{-1}$ in (a) and (b). In the normalized laser output
in (a) two features can be distinguished: On the one hand two strong
enhancement peaks between $t=-1$~ps and 4~ps and on the other hand an
oscillation around $I/I_0\approx 2$ for $t>4$~ps. The double peak structure
at the beginning looks very much like the dynamics in
Fig.~\ref{fig:rect_pulse}(a) and also the behavior of the inversion density
in (b) shows the above explained appearance of additional spectral holes at
higher energies. Thus we conclude that the features are due to the switch on
of the strain oscillation. The new feature is the oscillatory part for
$t>4$~ps, which has a period of half the strain period $\tau_{\rm
s}/2=\pi/\omega_{\rm s}\approx 3.14$~ps. The two red arrows mark maxima of
the intensity, which directly correspond to extrema of the strain
oscillation. The movement of the cavity resonance can be seen in (b) as
oscillating dark line around 0~meV. At every antinode of the strain shift the
cavity moves slower and the QDs have time to contribute to the laser process
more efficiently, which leads to the maxima in the intensity. Around the
nodes the QD emission can not follow properly and the intensity drops again.
In contrast to the pulsed excitation in Fig.~\ref{fig:rect_pulse}(a) the
intensity does not go back to $I/I_0=1$. This can easily be understood from
the dynamics of the outer spectral holes in Fig.~\ref{fig:rect_harm}(b). Like
in the pulsed case they decay after the cavity has moved out of resonance
with the respective QDs but gets back into resonance before the excitons are
fully re-occupied. Overall we find a broadening of the spectral hole for all
times after the strain field is switched on. This broadening spreads over the
full energy range that is covered by the strain shift of the cavity resonance
(or of the ensemble).

The situation shows some decisive differences when increasing the strain
frequency to $\omega_{\rm s}=10$~ps$^{-1}$ in Fig.~\ref{fig:rect_harm}(c) and
(d). The normalized intensity in (c) is still dominated by a pronounced
maximum lasting from $t=0$ to about 7~ps, which arises from the switch-on
process of the strain oscillation. At 15~ps the output has almost reached a
stationary value of about $I/I_0=2.5$. Still the whole curve is covered with
an oscillation with half the strain period $\tau_{\rm s}/2=\pi/\omega_{\rm
s}\approx 0.31$~ps, but with very small amplitudes compared to the previous
example. Also a very different picture is found for the inversion density in
(d). The oscillation of the cavity resonance in form of a dominating spectral
hole is no longer resolved. Within the first 5~ps multiple clearly separated
spectral holes appear symmetrically around $\hbar\omega-\hbar\omega_0=0$.
Apparently now a wide range of exciton energies contribute to the laser
process and lead to the significantly increased, almost constant, laser
intensity. A surprising feature is the appearance of spectral holes at
energies larger than the strain shift of $\eta_0^{\rm D}=10$~meV. The four
outer most lines in the plot are clearly above
$|\hbar\omega-\hbar\omega_0|=10$~meV. We will come back to this point in more
detail below.

The appearance of spectral holes at large energies becomes even more striking
when increasing the strain frequency further to $\omega_{\rm s}=50$~ps$^{-1}$
in Fig.~\ref{fig:rect_harm}(f). After the strain field is switched on the
inversion density reaches a stationary state after approximately 25~ps. Three
well separated spectral holes show up. While the one in the middle does not
change significantly from the unstrained case two additional ones appear at
approximately $\pm 30$~meV. For the laser intensity in (e) this enlarged
number of QDs contributing to the laser output leads to a slight increase.
Obviously the final intensity enhancement is smaller than in the case in (c).
Thus it can be expected that the strain frequency follows some sort of
resonance condition.

\begin{figure}
\centering
\includegraphics[width=0.8\columnwidth]{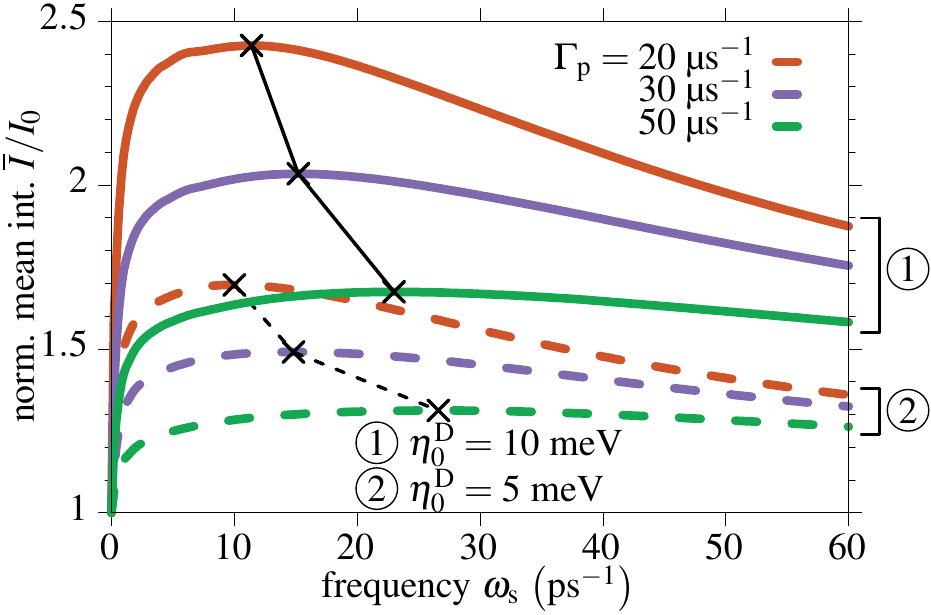}
\caption{Final mean enhancement for harmonic modulations of a flat ensemble as a function of the strain frequency $\omega_{\rm s}$. Different pump rates and strain amplitudes are given. The crosses mark the positions of the maxima.
\label{fig:rect_harm_enhance}
}
\end{figure}

\begin{figure*}
\centering
\includegraphics[width=0.8\textwidth]{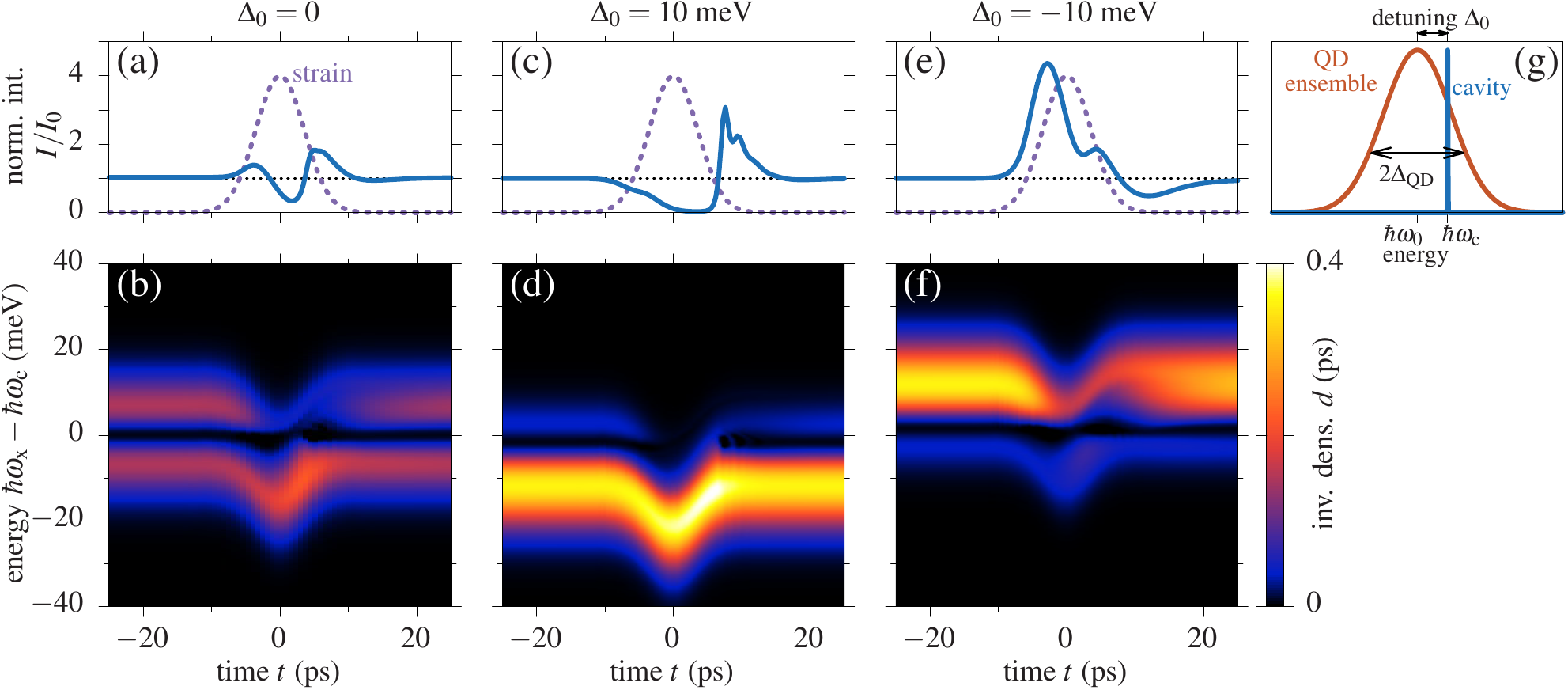}
\caption{Pulsed modulation of a Gaussian ensemble. The strain pulse is Gaussian with a energy shift of $\eta_0^{\rm D}=10$~meV and duration $\tau_{\rm s}=5$~ps. Top row: Normalized intensity $I/I_0$ as a function of time. The dashed violet curve shows the dynamics of the strain pulse. Bottom row: Inversion density $d$ as a function of time and energy for a fixed cavity resonance. (a), (b) Initial detuning $\Delta_0=0$; (c), (d) $\Delta_0=10$~meV; (e), (f) $\Delta_0=-10$~meV. (g) Schematic picture of the ensemble and the cavity mode.
\label{fig:gauss_puls}
}
\end{figure*}

Before coming back to the shape of the spectral hole under the strain
influence, we want to embark on the dependence of the enhancement on the
strain frequency. Figure~\ref{fig:rect_harm_enhance} shows the normalized
mean final intensity $\overline{I}/I_0$ as a function of the strain frequency
$\omega_{\rm s}$ for different strain amplitudes $\eta_0^{\rm D}$ and pump
rates $\Gamma_{\rm p}$ as listed in the picture. Every curve shows a more or
less pronounced maximum supporting the prediction, that the strain induced
enhancement works most efficiently for a characteristic frequency. For very
small strain frequencies the laser system can follow adiabatically and for
very fast oscillations the laser system is too inert and the response reduces
again. We also see that for higher strain amplitudes the enhancement factors
increase. This is intuitive because the strain-induced shift reaches a larger
amount of QDs if the amplitude of the shift is larger. With a close look at
the positions of the maxima (marked by the crosses in
Fig.~\ref{fig:rect_harm_enhance}) we find that the peak frequency shifts to
larger values for increasing pump rates. Thus the characteristic frequency of
the laser system becomes larger for growing pump rates. This growth of the
characteristic frequency can also be connected to the relaxation oscillations
during the switch-on of the laser. An example of the relaxation oscillations
in our system can be seen in Fig.~\ref{fig:gauss_puls}(c). It is well known
that also the frequency of the relaxation oscillations grows with increasing
pump power~\cite{bimberg2000qua}.

\begin{figure}[h]
\centering
\includegraphics[width=0.8\columnwidth]{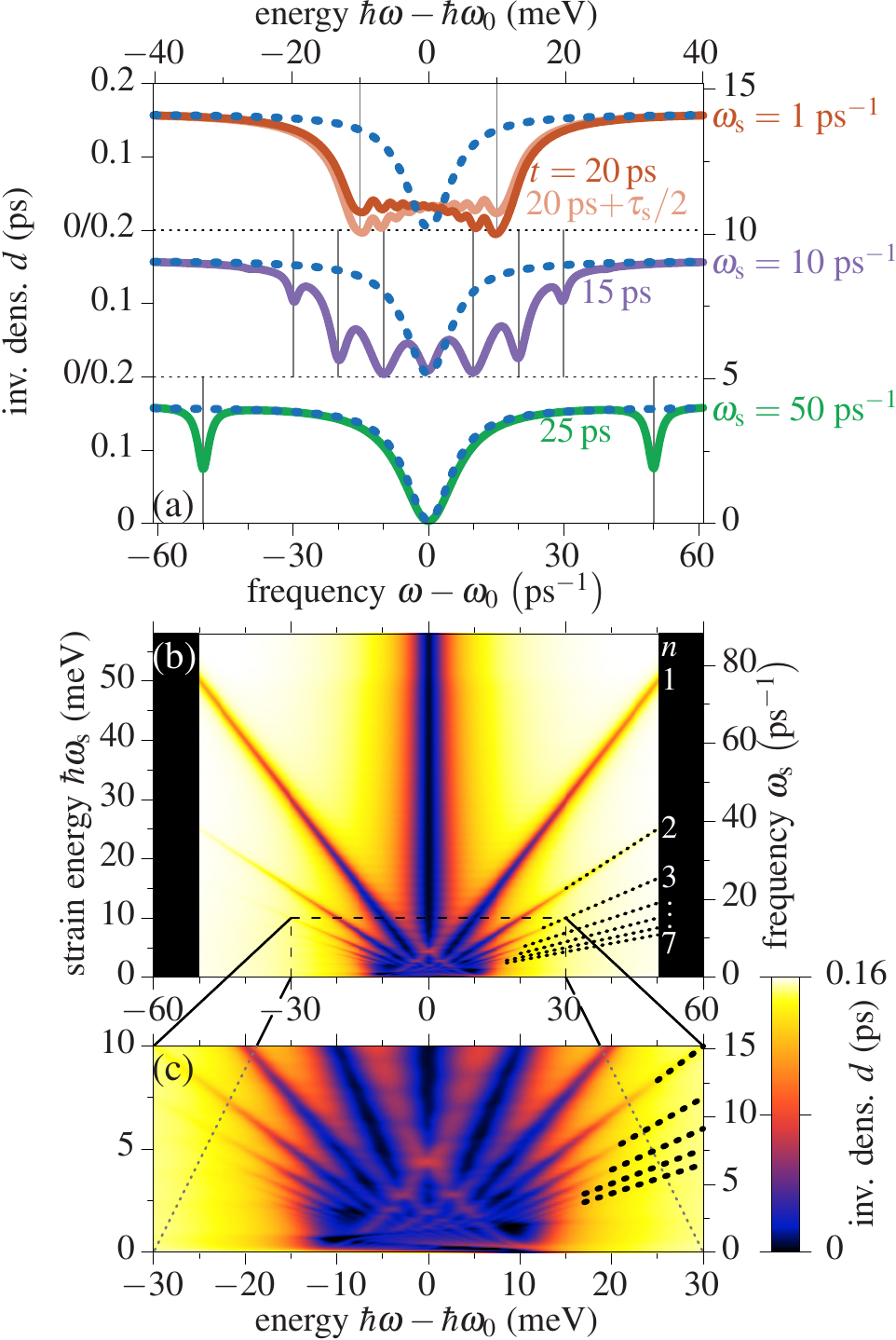}
\caption{Final spectral hole for harmonic modulations of a rectangular ensemble. (a) Final inversion density for the parameters in Fig.~\ref{fig:rect_harm} as a function of frequency $\omega$ (bottom axis) and energy $\hbar\omega$ (top axis). The dotted blue lines show the spectral hole without strain. (b), (c) Inversion density as a function of energy $\hbar\omega$ and strain energy $\hbar\omega_{\rm s}$ (left axis) or strain frequency $\omega_{\rm s}$ (right axis). (c) Zoom-in to the marked area in (b). The dotted black lines mark the position of the higher harmonics following Eq.~\eqref{eq:QDL_higher_harm} as listed on the right in (b).
\label{fig:rect_harm_spechole}
}
\end{figure}

The next focus is on the shape of the spectral hole and its dependence on the
strain frequency, which was already mentioned when discussing
Fig.~\ref{fig:rect_harm}. We have found that multiple depressions appear in
the inversion density. While for small strain frequencies these are
restricted to the strain shift amplitude, for increasing frequencies the
additional spectral holes also appeared at larger energies. To get a more
quantitative picture Fig.~\ref{fig:rect_harm_spechole}(a) shows the inversion
density at the end of the time windows shown in Fig.~\ref{fig:rect_harm}(b),
(d) and (f) from top to bottom, respectively. As a reference the spectral
hole without any strain is always plotted as dotted blue line. Note that the
bottom x-axis shows the exciton frequency $\omega-\omega_0$, while the top
one gives the energy $\hbar\omega-\hbar\omega_0$. For the case of
$\omega_{\rm s}=1$~ps$^{-1}$ (top, dark red) the spectral hole is very broad
with a rather flat base line, which has a slight negative tilt. That the
slope of the tilt oscillates with the strain frequency is confirmed by the
bright red curve, which shows the same inversion density but half a strain
period later. The complete hole spreads over the energy range, which is
covered by the strain shift ($\eta_0^{\rm D}=10$~meV) as can be seen from the
two vertical lines. The middle panel presents the $\omega_{\rm
s}=10$~ps$^{-1}$-case. Here we want to focus on the frequency axis at the
bottom. As marked by the vertical lines, the well separated minima in the
inversion density are exactly at multiples of the strain frequency, i.e., at
$\omega-\omega_0=\pm 10$, $\pm 20$ and $\pm30$~ps$^{-1}$. This is further
confirmed in the bottom panel for $\omega_{\rm s}=50$~ps$^{-1}$. Here the
additional spectral holes lie at $\pm 50$~ps$^{-1}$. The second one would
appear at $\pm 100$~ps$^{-1}$, which is already outside of the ensemble.

We can conclude that additional spectral holes appear at energies with
\begin{equation}
(\hbar\omega-\hbar\omega_0)_{\rm sh} = \pm n\hbar \omega_{\rm s}\,,\quad n\in \mathbb N_0\,,
\label{eq:QDL_higher_harm}
\end{equation}
i.e., at higher harmonics of the strain frequency.

To develop a complete picture of the strain frequency dependence of the
spectral hole Fig.~\ref{fig:rect_harm_spechole}(b) and (c) show the inversion
density of the QD ensemble at a fixed time long after the switch-on process
of the strain field as a function of the exciton energy
$\hbar\omega-\hbar\omega_0$ and the strain energy according to its frequency
$\hbar\omega_{\rm s}$. Panel (c) is a zoom-in on the dashed area in (b).
Focussing first on the large energy range in (b) we see that the central
spectral hole at the initial position of the cavity is almost unperturbed by
the strain field. For large strain energies the additional spectral holes
show up as diagonal lines that decay for increasing energies. The dotted
black lines follow Eq.\eqref{eq:QDL_higher_harm} together with the number of the
higher harmonics $n$ given on the right. Within the Floquet theory it is
shown that a harmonically driven two level system with the energy splitting
$\hbar\Omega$ has multiple additional so called \emph{Floquet states} at
energies $\hbar\Omega\pm n \hbar\omega_{\rm d}$, where $\omega_{\rm d}$ is
the driving frequency. Therefore QDs that are detuned from the cavity mode by
multiples of the strain frequency gain new transitions that are in resonance
with the cavity. By this they can contribute to the laser process. A more
intuitive interpretation of the addition transitions is as follows: Because
the QD ensemble is interacting with a phonon system with a discrete spectrum,
single or multiple absorptions and emissions of phonons are possible to
overcome the detuning to the cavity mode. Accordingly, the outer spectral
holes arise from excitons, which contribute to the laser process via phonon
assisted transitions. Focussing on the region of small energies in (c) we
find a transition area from the adiabatic situation ($\hbar\omega_{\rm
s}\approx 0$~meV), where the spectral hole at the position of the cavity mode
can immediately follow the shift of the ensemble, toward the formation of the
additional higher harmonic spectral holes for strain energies
$\hbar\omega>2.5$~meV. Below this energy the widening of the spectral hole is
restricted to the range of energies that is covered by the strain amplitude
$\eta_0^{\rm D}$, i.e., $\pm 10$~meV. Below $\hbar\omega=12.5$~meV also the
central spectral hole is not well developed and vanishes within the whole broadening.
%

\subsection{Gaussian ensemble}
\label{sec:gauss}
In the second example, we now want to focus on the adiabatic shift. For this
purpose we choose a QD ensemble shape with a Gaussian distribution given by
\begin{equation}
n(\omega)= \frac{\hbar N_{\rm QD}}{\sqrt{\pi}\Delta_{\rm QD}} \exp\left[-\left(\frac{\hbar\omega-\hbar\omega_0}{\Delta_{\rm QD}}\right)^2\right]\,,
\end{equation}
which is the same as in Ref.~\cite{czerniuk2017pic}. Any energetic
shift of the ensemble leads to a variation of the number of QDs in resonance
with the cavity mode, therefore the adiabatic shift of the ensemble does
inevitable play an important role. On the other hand the shaking effect cannot be suppressed by a specific choice of the ensemble and will always
contribute. A schematic picture of the ensemble is given in
Fig.~\ref{fig:gauss_puls}(g), which also shows the initial detuning
\begin{equation}
\Delta_0 = \hbar\omega_{\rm c}-\hbar\omega_0
\end{equation}
between the cavity mode and the ensemble center. The total number of QDs is
again $N_{\rm QD}=5\times 10^4$.
%
\subsubsection{Single strain pulse}
As a starting point we choose a monopolar strain pulse to interact with the system. Such pulses have also an experimental counterpart, namely acoustic solitons, which develop during the nonlinear propagation of the phonons. They are of special interest, since they combine high frequency phonons with a very small spatial extension of only a few nanometers. The dynamics is given by a single Gaussian of the form
\begin{equation}
D \eta(t) = -\eta^{\rm D}_0\,  \exp\left[ - \left(\frac{t}{\tau_{\rm s}}\right)^2\right]\ .
\label{eq:QDLstrain_puls_gauss}
\end{equation}
The amplitude of the strain shift is $\eta_0^{\rm D}$ and the width is given
by $\tau_{\rm s}$. This strain dynamics is a good approximation for soliton
pulses, which develop during the nonlinear propagation of acoustic
pulses~\cite{vancapel2015non} [cf. Fig.~\ref{fig:motiv}(a)]. This choice of
the strain pulse is even simpler than the bipolar pulse in
Eq.~\eqref{eq:strain_pulse}, in order to demonstrate the effect of the
adiabatic ensemble shift more transparently. Figure~\ref{fig:gauss_harm}
shows the results for an ensemble width of $\Delta_{\rm QD}=10$~meV, a pump
rate of $\Gamma_{\rm p}=10$~ps$^{-1}$, a strain amplitude of $\eta_0^{\rm
D}=10$~meV and a pulse duration of $\tau_{\rm s}=5$~ps.

We start the discussion with the special case of vanishing initial detuning
$\Delta_0=0$. Due to the symmetry of the system the reaction on the strain
pulse does not depend on the sign of the shift. The results for the
normalized laser intensity are shown in Fig.~\ref{fig:gauss_puls}(a) (solid
blue) together with the strain pulse (dotted violet). We see that the laser
output is first increased, then decreased and increased another time during
the time of the strain pulse. We directly compare these dynamics with the
behavior of the inversion density in (b), which is here shown for a fixed
cavity resonance, i.e., against the energy $\hbar\omega_{\rm
x}-\hbar\omega_{\rm c}$. Thus in the figure the ensemble is shifted by the
strain field. Before the strain pulse we clearly see the spectral hole in the
center of the Gaussian QD distribution, during the pulse the ensemble shifts
to smaller energies and therefore out of resonance with the cavity. Without
the shaking effect, we would expect a drop of the laser intensity, because
the number of QDs in resonance with the laser mode is reduced during the
first half of the strain pulse. However, we contrary first observe an
increase of the intensity in (a), which proves that the shaking effect
outweighs the adiabatic response due to the reduced number of resonant QDs.
Likewise the attenuation of the output can be understood because it happens
at the maximum of the strain shift, which happens much slower and the
adiabatic shift becomes more important. During the second half of the strain
pulse the ensemble is brought back into resonance with the cavity, such that
the shaking effect and the adiabatic shift work in the same direction and
result in a slightly stronger enhancement in (a). As can be seen in (b)
around $t=0$ the inversion density increases while the ensemble is out of
resonance with the cavity. When these re-occupied dots get back into the
cavity mode they can emit and contribute to the laser process.

In the other two examples the initial detuning is chosen in the same order as
the strain amplitude. In (c) and (d) the initial detuning is
$\Delta_0=10$~meV, but the strain shifts the ensemble further away from
cavity. For the laser intensity in (c) this means a complete quenching at
$t=0$. During the second half of the strain pulse the laser is switched on
again exhibiting the characteristic relaxation oscillations before reaching
its original intensity. In the inversion density in (d) we see that the
spectral hole is now at the edge of the ensemble. During the strain pulse the
ensemble shifts to even smaller energies and the spectral hole in the cavity
resonance vanishes, leaving an almost completely occupied Gaussian
ensemble. When the ensemble moves back into resonance with the cavity also
the dynamics of the spectral hole reveal  the relaxation oscillations. To
conclude, the strain leads to a switch-off of the laser due to the adiabatic
reduction of resonant QDs. The pronounced enhancement of the output happens
due to the switch-on process of the laser and is part of the characteristic
relaxation oscillations.

In Fig.~\ref{fig:gauss_puls}(e) and (f) the situation is inverted, the
initial detuning is chosen to $\Delta_0=-10$~meV and the strain pulse shifts
ensemble and cavity into resonance at its maximum. In the intensity in (e)
this leads to a strong enhancement peak. At the end of the pulse around
$t=10$~ps we also find a significant reduction of the output. Looking at the
inversion density in (f) we see that initially the main part of the ensemble
is on the other side of the cavity. The strain pulse then reduces the
detuning. It is clearly visible that most of the QDs become de-occupied
around $t=0$. When this large number of QDs emits stimulatedly into the
cavity mode, the large enhancement in (e) is observed. Also while the
ensemble is shifted back to the original detuning situation at $t=10$~ps, the
laser field is still enhanced. On the other hand due to the previous
stimulated emissions there is a reduced number of occupied excitons, which are
not capable of feeding the laser process. Therefore the intensity drops below
$I_0$, while enough dots recover their occupation. The system reaches its
initial stationary value after $t\approx25$~ps.
%

\subsubsection{Harmonic modulation}
\begin{figure}
\centering
\includegraphics[width=0.8\columnwidth]{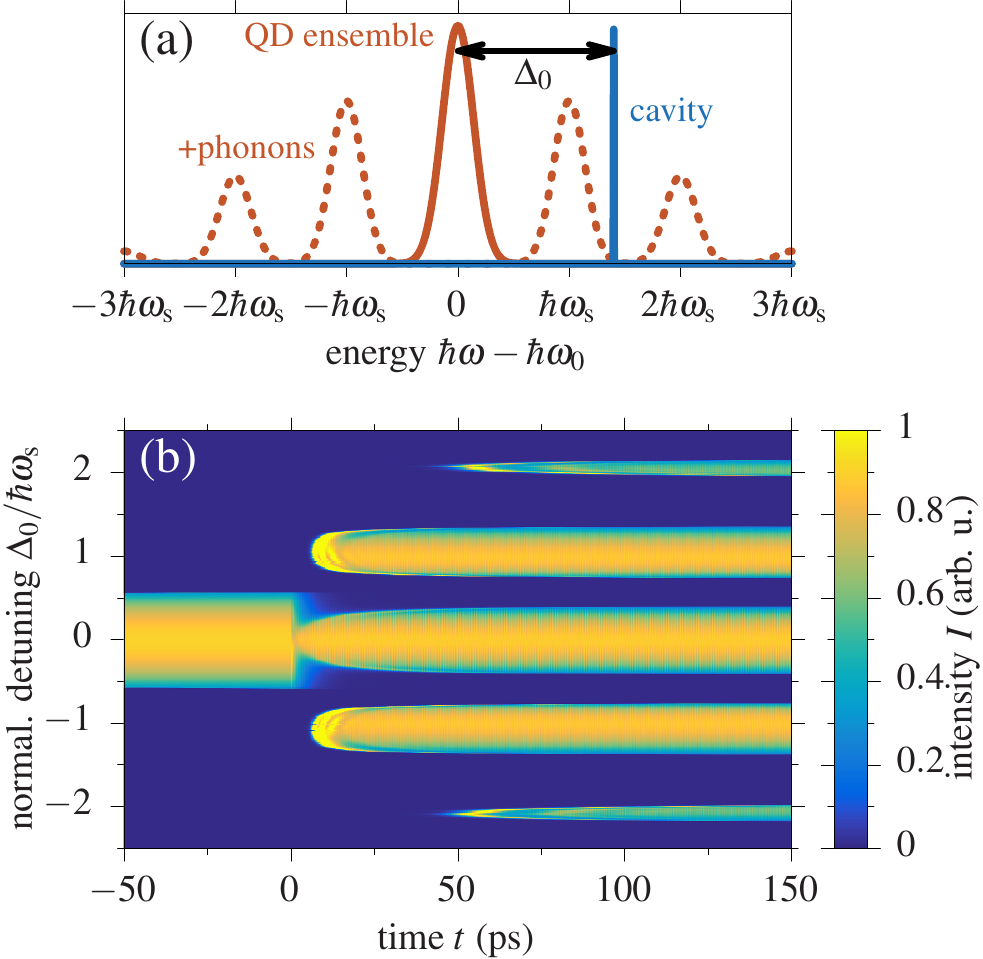}
\caption{Harmonic modulation of a Gaussian ensemble. (a)~Schematic picture of the QD density of a Gaussian ensemble with harmonic strain driving. Phonon assisted satellite peaks are given as dashed line. (b) Laser intensity $I$ as a function of time and initial detuning $\Delta_0$ normalized to the strain energy $\hbar\omega_{\rm s}$.
\label{fig:gauss_harm}
}
\end{figure}

In this part we come back to the switched on harmonic strain oscillation from Eq.\eqref{eq:QDL_strain_harm} and apply it to a Gaussian ensemble. We choose a
small ensemble width of $\Delta_{\rm QD}=2$~meV, a strain frequency of
$\omega_{\rm s}=30$~ps$^{-1}$ and a strain amplitude of $\eta_0^{\rm
D}=25$~meV. Note that these parameters were chosen to optimize the visibility
of the effect. Figure~\ref{fig:gauss_harm}(a) shows a schematic picture of
the QD ensemble (solid orange) together with the cavity mode (blue).
Following the arguments of the Floquet theory in Sec.~\ref{sec:rect} the
interaction with the single frequency phonons of the strain field will lead
to additional satellite replicas of the ensemble shifted by multiples of the
strain energy $\hbar\omega_{\rm s}$, which are indicated by the dotted orange
curve. This effect has been seen in photoluminescence spectra of single QDs
that were driven by surface acoustic waves~\cite{metcalfe2010res}.

In Fig.~\ref{fig:gauss_harm}(b) we see the laser intensity as a function of
time $t$ and detuning $\Delta_0$ normalized with respect to the strain energy
$\hbar\omega_{\rm s}$. Before the switch-on of the strain oscillation, i.e.,
for $t<0$~ps, we find an approximately Gaussian profile of the emission
intensity, which directly resembles the ensemble shape because the laser
intensity increases with the number of resonant QDs. When the strain field is
switched on additional yellow lines appear at multiples of the strain energy.
This shows that the laser system also works in a completely phonon assisted
situation. A close look at the intensity amplitudes of the assisted lines
shows that they are less strong than the central resonant one. When the
strain field is switched on at $t=0$ also the central line is significantly
narrowed due to the interaction with the phonons, but recovers almost
completely after 150~ps. The lines at $\Delta_0/\hbar\omega_{\rm s}=\pm 2$
appear later, because it takes more time for the system to start and reach
the final output. Here it is more clearly visible that their intensity is
less strong, reaching only half the resonant amplitude.

%
%
\section{Comparison to experiment}
\label{sec:compare}
With our newly gained knowledge we close the discussion by shortly revisiting
the experimental results in Fig.~\ref{fig:motiv}. Let us first comment on the
complex strain pulse, which is obtained in the experiment. Its dynamics
arises from the propagation of the phonons through the sample before reaching
the active medium. The injection of the acoustic pulse happens via an intense
picosecond laser pulse that rapidly heats an aluminum film as shown on the
left side of Fig.~\ref{fig:motiv}(a). The heating is associated with an
expansion of the film, which launches a phonon wave packet with an
approximately Gaussian lattice displacement into the substrate material. This
lattice displacement corresponds to a bipolar strain pulse of about 20~ps.
Due to the large strain amplitudes on the order of $\eta\approx 10^{-3}$
nonlinear effects have to be taken into account in the phonon propagation
through the substrate~\cite{scherbakov2007chi,vancapel2015non}. During this
propagation solitons form and the strain pulse gets stretched over a few tens
of picoseconds. Before reaching the QD layer of the sample the strain pulse
is additionally modified by passing through the bottom DBR [left one in
Fig.~\ref{fig:motiv}(a)]. Multiple transmission and reflection processes in
the DBR then lead to the final dynamics in Fig.~\ref{fig:motiv}(b) that lasts
several hundreds of picoseconds.

In the laser output [Fig.~\ref{fig:motiv}(c)] we find first a strong
enhancement peak followed by 100~ps of almost complete quenching. After
200~ps the output is dominated by a harmonic oscillation, reproduced well by
the simulation [Fig.~\ref{fig:motiv}(d)]. In the presented example the cavity
mode is positively detuned from the QD ensemble, such that negative strain
values reduce the detuning and positive ones increase it. Due to our previous
analysis we are now able to attribute these variations to the two effects:
Through the first negative solitonic strain pulses both -- shaking and
shifting -- lead to an increase of the laser output as it was found in
Fig.~\ref{fig:gauss_puls}(e). The following attenuation then has two reasons.
On the one hand many QDs were de-occupied during the first intensity peak and
the initial intensity cannot be preserved any longer. On the other hand the
mean strain value is positive, which reduced the laser output naturally
through the adiabatic shift. The harmonic oscillation in the last part in
Fig.~\ref{fig:motiv}(b) arises from the adiabatic shift due to the slow
harmonic oscillation of the strain field.
%
\section{Conclusion}
\label{sec:concl}
In summary we have presented a systematic study of the QD laser emission
properties driven by coherent phonons. By choosing a rectangular ensemble the
adiabatic shift of the ensemble could be suppressed and the shaking effect
was analyzed separately. We have shown that the intensity enhancement depends
strongly on the time scale of the phonon field. For harmonic strain
oscillations phonon assisted transitions contribute to the laser output. For
a realistic Gaussian ensemble both effects take part in the process. We
have found that depending on the initial detuning, the intensity could either
be significantly enhanced or switched-off on an ultra fast time scale.

The detailed study of the different effects allowed us to obtain a deep
understanding of the details of the measured intensity modulations. With
this, it will be possible to tailor the nanostructure as well as the phonon
pulses to reach a desired control of the laser output, may it be a long
lasting amplification or a brief enhancement with a strong amplitude. Hence,
our work paves the way for further research both on the theoretical side,
e.g. for further optimization, and on the experimental side, e.g. to
fabricate and measure different structures and phonon pulses. 

The demonstrated control of light-matter interaction by acoustic waves may be extended also into the quantum regime, for example for the deterministic generation of single or entangled photons from a quantum dot in an optical microresonator. This can be achieved by shifting the exciton transition dynamically to the optical cavity mode so that a single photon is emitted in resonance condition. Similarly the two transitions in the bixeciton radiative decade maybe sequentially shifted into cavity resonance. Thereby not only true exciton-photon control may be obtained but also the temporal jittering of the photon emission times may be reduced.
%
\section{Acknowledgements}
The Dortmund team acknowledges the financial support by the Collaborative Research Center TRR 142. We are grateful for fruitful discussions with Andrey V. Akimov.
%
%

\end{document}